\documentclass[titlepage,revtex,12pt]{article}
\usepackage{amsmath,amssymb}
\begin{document}

\begin{titlepage}
\thispagestyle{empty}
\begin{center}
{\Large\bf Packet Spreading and Einstein Retardation}\\[0.5cm]
{\large\bf M. I. Shirokov }\\[0.5cm]
{Bogoliubov Laboratory of Theoretical Physics\\
Joint Institute for Nuclear Research\\
141980 Dubna, Russia\\
e-mail: shirokov@theor.jinr.ru}
\end{center}

\vspace*{1cm} \noindent
{\large Abstract}\\

According to the classical special theory of relativity any nonstationary
system moving with velocity $v$ must evolve (e.g., decay) $1/\gamma$ times
slower than the system at rest, $\gamma =(1-v^2)^{-1/2}$ (the Einstein
retardation ER). Quantum mechanics allows one to calculate the evolution of
both systems separately and to compare them thus verifying ER.
It is shown here that ER is not valid for a simple system: the spreading packet
of the free spinless particle. Earlier it was shown that ER does not hold
for some other systems. So one may state that ER is not a universal
kinematic law in quantum mechanics.
\end{titlepage}


\section{INTRODUCTION}
\label{intro}

Experiments show that moving unstable particles (e.g., $\mu$-mesons, $\pi$-mesons)
decay slower than the particles at rest. More explicitly, let
$N_0(t)=N\exp (-t/\tau_0)$ be the decay law of the particles at rest,
$\tau_0$ being life-time. Then the life time of particles moving with velocity
$\vec v$ is $\tau_v=\tau_0\gamma$, $\gamma =(1-v^2)^{-1/2}$, and the decay law
of moving particles is
\begin{equation}
\label{eq1}
N_v(t)=N\exp (-t/\tau_v)=N\exp (-t/\tau_0\gamma)=N_0(t/\gamma).
\end{equation}
One may rewrite Eq. (\ref{eq1}) as
\begin{equation}
\label{eq2}
N_p(t)=N_0(t/\gamma), \qquad \gamma =\sqrt{p^2+m^2}/m
\end{equation}
using the corresponding momentum $\vec p$ instead of velocity $\vec v$:
$\vec p=E\vec v$.

The usual theoretical explanation of Eqs. (\ref{eq1}), (\ref{eq2}) is
based on the Einsteinian special relativity theory. It is set forth as follows.
A moving clock has a slower course as compared with the clock at rest, namely
$\gamma d\tau =dt$, e.g., see \cite{c1}, Ch.~2, Eqs.~(36) or (38).
The unstable substance may serve as a clock, see \cite{c1}, Ch.~2. Being the clock,
the moving ensemble of unstable particles must decay slower than the
ensemble at rest. This is described by Eq. (\ref{eq1}): $N_v$ assumes at the
moment $t$ the value which $N_0$ assumes at the moment $t/\gamma$.

This argumentation may be applied to any nonstationary physical system which
may serve as a clock. Instead of $N_v(t)$ another time-dependent observable
$F_v(t)$ may be considered. As the example, the dispersion $\sigma^2(t)$ of
the spreading packet may be examined, see Sect.~\ref{s2} below. In the same manner
as above one may argue that the equation
\begin{equation}
\label{eq3}
F_v(t)=F_0(t/\gamma)
\end{equation}
must hold. Equation (\ref{eq1}) is a particular case of Eq. (\ref{eq3}).
Eq. (\ref{eq3}) means that $F_v$ assumes at the moment $t$ the value which $F_0$
assumes at the earlier moment $t/\gamma$. I call relation (\ref{eq3}) Einsteinian
retardation ER. It is a kinematic law in special relativity, see \cite{c1}, Ch.~2.

However, clocks considered in special relativity are nonquantum objects:
they have simultaneously a definite position (e.g., being in frame's origin)
and definite velocity (e.g., zero velocity), see the beginning of
Ch.~(2.6) in \cite{c1}. This is impossible for such quantum objects as $\mu$
or $\pi$ mesons. Therefore, the usual explanation of relations (\ref{eq1}),
(\ref{eq2}), (\ref{eq3}) is not valid for quantum objects.

However, one may verify the validity of these relations in quantum mechanics.
The number of particles and other time-dependent observables may be considered
as quantum observables. Using quantum mechanics one may calculate the observables
separately for the moving system and the system at rest. Comparing them one may
ascertain whether Eqs. (\ref{eq1}), (\ref{eq2}), (\ref{eq3}) hold. For unstable
particles this approach was considered in \cite{c2}-\cite{c6}. The result may be formulated
as follows: ER does not hold exactly but it is valid up to high precision.

Oscillating systems ($K_0$-$\bar{K_0}$ mesons and oscillating neutrino) were
considered in \cite{c5}, \cite{c6}. It is shown that large deviations from ER may exist.

A moving nonstationary system was discussed in refs.~\cite{c6}, \cite{c7} which evolves
\underline{faster} than the system at rest: $F_v(t)=F_0(\gamma t)$ holds
instead of Eq. (\ref{eq3})!

In this paper I consider in Sect.~\ref{s2} the simple nonstationary system:
the spreading wave packet of the free spinless particle. Packet dispersions
(see below Eq. (\ref{eq4})) are used as the time-dependent observables which
describe packet spreading. In Sect.~\ref{s2} I calculate the longitudinal
dispersion $\sigma_l^2(\vec{v},t)$ (dispersion along the packet velocity
$\vec v=\vec p/E$). It is compared in Sect.~\ref{s3} with the dispersion
$\sigma^2(0,t)$ of the packet at rest. The connection
$\sigma_l^2(\vec{v}, t)=\sigma^2(0,t/\gamma^3)$ is obtained. So
$\sigma_l^2(\vec{v},t)$ is retarded as compared to $\sigma^2(0,t)$
but Eq. (\ref{eq3}) is not valid, i.e. ER fails. The premises of
the result are summed up in Sect.~\ref{s3}.


\section{Dispersions of Gaussian packet}
\label{s2}

ER is the kinematic statement on the time evolution of nonstationary
physical systems which may be considered as clocks. So the quantum mechanical
consideration of ER must deal with time-dependent observables (so that the
known $S$-matrix approach is not relevant).

In the capacity of the nonstationary system let us consider the spreading packet
of the scalar particle. Let us consider the packet dispersions
$\sigma^2_1(t)$, $\sigma^2_2(t)$, $\sigma^2_3(t)$
\begin{equation}
\label{eq4}
\sigma^2_j(t)=\int d^3x x^2_j \rho (\vec x,t)
-\left[ \int d^3x x_j \rho (\vec x,t) \right]^2,
\quad j=1,\,2,\,3
\end{equation}
as time-dependent observables. Here $\rho (\vec x,t)$ is the probability density
to find the particle at the point $\vec x$ at time $t$. The density and dispersions
may be experimentally measured. In quantum theory $\rho$ is expressed in terms of
the packet wave function $\Psi$, the positive-energy solution of the Klein-Gordon
equation
\begin{equation}
\label{eq5}
i\partial\Psi /\partial t=\hat{E}\Psi, \quad
\hat{E}\equiv\left[ (-i\partial /\partial\vec x)^2+m^2\right]^{1/2}.
\end{equation}
The known usual expression of $\rho$ is
$\rho (\vec x,t)\sim\hat{E}\Psi^*(\vec x,t)\Psi (\vec x,t)
+\Psi^*(\vec x,t)\hat{E}\Psi (\vec x,t)$, e.g. see \cite{c8}, Ch.~3.
However, the expression is not positive definite function of $\vec x$,
e.g. see \cite{c9}, Supplement~II. Therefore, it does not suit as
a probability density, although $\int\rho (\vec x,t)d^3x$ is positive and may be
normalized to unity.

Here I use Newton-Wigner wave function $\Psi_{\rm NW}$, see \cite{c10}, Eq.~(5).
In their representation $\rho (\vec x,t)=\Psi_{\rm NW}^*(\vec x,t)\Psi_{\rm NW} (\vec x,t)$,
see \cite{c10}, Eq.~(6). In this equation and in what follows the letter $\vec x$
denotes the Newton-Wigner coordinate, see \cite{c8}, Ch.~3. The function $\Psi_{\rm NW}$
will be denoted by $\Psi$.

The solution of (\ref{eq5}) then may be represented as
\begin{equation}
\label{eq6}
\Psi (\vec x,t)=(2\pi)^{-3/2}\!\!\!\int\!\! d^3k \exp (i\vec k\vec x)
\Phi (\vec k) \exp (-itE_k), \quad\!\!\! E_k=\sqrt{k^2+m^2},
\end{equation}
see \cite{c8}, Chs.~7 and 3; \cite{c10}, Eq.~(5). $\Phi (\vec k)$ is the initial
wave function of the packet in momentum representation.
%
For $\Phi (\vec k)$ let us choose the product of three Gaussian packets
\begin{equation}
\label{eq7}
\Phi (\vec k)=\varphi_1(k_1) \varphi_2(k_2) \varphi_3(k_3),
\quad \varphi_j(k_j)=M\exp [-(k_j-p_j)^2\sigma^2].
\end{equation}
The functions $\varphi_j$, $j=1,\,2,\,3$, are normalized to unity
\begin{equation}
\label{eq8}
\int dk_j |\varphi_j(k_j)|^2 =1
\end{equation}
if $M^2=\sigma\sqrt{2/\pi}$. Then $\Phi (\vec k)$ is also normalized:
$\int d^3k |\Phi(\vec k)|^2 =1$.

It is easy to show that the parameters $p_j$ in Eq. (\ref{eq7}) are
components of the mean momentum of the packet:
\begin{eqnarray}
\label{eq9}
\int d^3k k_j |\Phi(\vec k)|^2&=&M^2\int_{-\infty}^{+\infty} dk_j
k_j |\varphi_j(k_j)|^2 \nonumber\\
&=& M^2\int dk'_j (p_j+k'_j)\exp [-2(k')^2\sigma^2]=p_j .
\end{eqnarray}
The derivation uses the normalization (\ref{eq8}), the change of the integration
variables $k'_j=k_j-p_j$, the parity of the function $\exp [-2(k')^2\sigma^2]$.

The initial wave function in the coordinate representation $\Psi (\vec x,0)$,
see Eq. (\ref{eq6}) at $t=0$, also reduces to the product of three factors
$$\Psi (\vec x,0)=\psi(x_1)\psi(x_2)\psi(x_3).$$
However $\Psi (\vec x,t)$, Eq. (\ref{eq5}), cannot be represented in such a simple
form because of the factor
$$\exp [-it(k_1^2+k_2^2+k_3^2+m^2)^{1/2}]$$
in the integrand of Eq. (\ref{eq6}).

The triple integral in Eq. (\ref{eq6}) may be calculated approximately if
the parameter $\sigma$ is large enough, e.g. cf. \cite{c11}, Ch.~3. To show this,
let us change the integration variables $\vec k'=\vec p - \vec k$
in Eq. (\ref{eq6}):
\begin{eqnarray}
\label{eq10}
\Psi (\vec x,t)&=&(2\pi)^{-3/2}M^3\int d^3k' \exp [i(\vec p - \vec k')\vec x]
\nonumber\\
&&\times\exp [-(\vec k')^2\sigma^2] \exp\{ -it[(\vec p - \vec k')^2+m^2]^{1/2}\}.
\end{eqnarray}
The function $\exp [-(\vec k')^2\sigma^2]$ cuts off the values of $(\vec k')^2$
which are much larger than $1/\sigma^2$. So one may assume, e.g., $k'<3/\sigma$.
Let $\sigma$ be much larger than the Compton wave length $\lambda_m=1/m$,
e.g. $\sigma >3\lambda_m$ or $3/\sigma <m$. It will be shown below that
$\sigma^2$ is space dispersion of the initial packet (see Eq. (\ref{eq24})).
It follows from the inequalities $k'<3/\sigma$ and $3/\sigma <m$ that
$k'\ll m$. Then $k'\ll\sqrt{p^2+m^2}\equiv E$ all the more. As $k'/E\ll 1$,
one may expand
\begin{eqnarray}
\sqrt{(\vec p - \vec k')^2+m^2}&=&
\sqrt{\vec p^2+m^2+(\vec k')^2-2(\vec p \vec k')}
\nonumber\\
&=& E\sqrt{1-2(\vec p \vec k')/E^2+(\vec k')^2/E^2} \nonumber
\end{eqnarray}
in the series over degrees of $k'/E$. Using the expansion
$$\sqrt{1+\alpha}=1+\alpha /2-\alpha^2/8+\ldots , \quad
\alpha =-2(\vec p \vec k')/E^2+(\vec k')^2/E^2$$
and neglecting the term smaller than $(k'/E)^2$ one gets
\begin{equation}
\label{eq11}
\sqrt{(\vec p - \vec k')^2\!+m^2}\cong E[1-(\vec p \vec k')/E^2
+(\vec k')^2/2E^2-(\vec v \vec k')/2E^2], \, \vec v=\vec p/E.
\end{equation}
Let us direct the third axis $\vec e_3$ ($\vec e_z$) of the coordinate frame
along $\vec p$ so that $\vec p=(0,0,p)$ and $\vec v=(0,0,v)$
($p$ denotes $|\vec p|$ and $v$ denotes $|\vec v|$).
In this frame (\ref{eq11}) turns into
\begin{equation}
\label{eq12}
\sqrt{(\vec p - \vec k')^2\!+m^2}\cong E\{ 1-pk'_z/E^2+[(k'_1)^2
+(k'_2)^2+(1-v^2)(k'_3)^2]/2E^2\}.
\end{equation}
Note that no supposition on $p$ value has been assumed so that
$0\leq p<\infty$. Using the approximation (\ref{eq12}) in Eq. (\ref{eq10})
one gets that the triple integral in Eq. (\ref{eq10}) reduces to the product
of three single-valued integrals:
$$\Psi (\vec x,t)\cong\psi_1(x_1,t)\psi_2(x_2,t)\psi_3(x_3,t),$$
\begin{eqnarray}
\label{eq13}
\psi_1(x_1,t)&=&(2\pi)^{-1/2}MI_1(x_1,t);\,
\psi_2(x_2,t)=(2\pi)^{-1/2}MI_2(x_2,t);\nonumber\\
\\
\label{eq14}
\psi_3(x_3,t)&=&(2\pi)^{-1/2}M\exp [ipx_3-itE]I_3(x_3,t);\\
\label{eq15}
I_j(x_j,t)&=&\int_{-\infty}^{+\infty}dk'_j \exp [-ik'_j(x_j-v_jt)]
\exp [-i(k'_j)^2(a_j)^2], \nonumber\\
&&j=1,\,2,\,3;\\
\label{eq16}
a_1^2=a_2^2&=&\sigma^2+it/2E, \quad a_3^2=\sigma^2+it(1-v^2)/2E,
\nonumber\\
v_1=v_2&=& 0, \quad v_3=v.
\end{eqnarray}
For integrals $I_j(x_j,t)$, Eq. (\ref{eq15}), see e.g. \cite{c12}, Ch.~2.5.36.1:
\begin{equation}
\label{eq17}
I_j(x_j,t)=\sqrt{\pi/a_j}\exp [-(x_j-v_jt)^2/4a_j^2].
\end{equation}
Using other tabular integrals one may verify that $\psi_j(x_j,t)$,
Eqs. (\ref{eq13}), (\ref{eq14}), (\ref{eq15}), are normalized:
\begin{equation}
\label{eq18}
\int dx_j|\psi_j(x_j,t)|^2=(2\pi)^{-1}M^2
\int_{-\infty}^{+\infty}dx_j|I_j(x_j,t)|^2=1.
\end{equation}

Let us calculate the mean positions $X_n(t)$, $n=1,\,2,\,3$, of the moving
packet at the moment $t$:
\begin{eqnarray}
\label{eq19}
X_n(t)&=&\int\!\int\!\int d^3x x_n|\Psi (\vec x,t)|^2\cong
\int dx_n x_n|\psi_n (x_n,t)|^2 \nonumber\\
&=& (2\pi)^{-1}M^2\int dx_n x_n|I_n (x_n,t)|^2.
\end{eqnarray}
Here $I$ use the normalization (\ref{eq18}) of the function $\psi_j$
with $j\neq n$ and then use Eqs. (\ref{eq13}), (\ref{eq14}).
Further the change $x'_n=x_n-v_nt$ of the integration variables is used
in the last integral in Eq. (\ref{eq19}). Finally, the parity of $I_n$ is
taken into account: $I_n (x'_n,t)=I_n (-x'_n,t)$.
One obtains the result
\begin{equation}
\label{eq20}
X_1(t)\cong X_2(t)\cong 0, \quad\!\!\! X_3\cong vt,
\quad\!\!\! v=|\vec v|=v_3=p/E, \quad\!\!\! E=\sqrt{p^2+m^2}.
\end{equation}
This means that the center of the packet moves along $\vec p$ with
the velocity $\vec v=\vec p/E$. In addition, the packet spreads,
the spreading being characterized by the packet dispersions
$\sigma_1^2$, $\sigma_2^2$, $\sigma_3^2$, see Eq. (\ref{eq4}).

Let us name $\sigma_3^2(\vec p,t)$ the longitudinal dispersion
$\sigma_l^2(\vec p,t)$ (dispersion along $\vec p$) and
$\sigma_1^2$, $\sigma_2^2$ transversal ones. Using
Eqs. (\ref{eq4}), (\ref{eq19}), (\ref{eq20}) one obtains for $\sigma_l^2$:
\begin{eqnarray}
\label{eq21}
\sigma_l^2(\vec p,t)&\equiv &\sigma_3^2(\vec p,t)=
\int dx_3 x_3^2|\psi_3 (x_3,t)|^2 -(vt)^2 \nonumber\\
&=& (2\pi)^{-1}M^2\int dx_3 x_3^2|I_3 (x_3,t)|^2 -(vt)^2.
\end{eqnarray}
The further derivation of $\sigma_l^2$ is more tedious than the
calculation of $X_3(t)$. The result is
\begin{equation}
\label{eq22}
\sigma_l^2(\vec p,t)=\sigma^2+t^2(1-v^2)^2/4E^2\sigma^2,
\quad E=\sqrt{p^2+m^2}=m\gamma.
\end{equation}
In the same manner one obtains the transversal dispersions
\begin{equation}
\label{eq23}
\sigma_1^2(\vec p,t)=\sigma_2^2(\vec p,t)=\sigma^2+t^2/4E^2\sigma^2.
\end{equation}
If the packet is at rest ($v=0$, $p=0$, $E=m$) all dispersions are equal:
\begin{equation}
\label{eq24}
\sigma_l^2(0,t)=\sigma_1^2(0,t)=\sigma_2^2(0,t)=\sigma^2+t^2/4m^2\sigma^2.
\end{equation}
When $t=0$ Eqs. (\ref{eq21}), (\ref{eq22}) turn into the initial dispersions:
$$\sigma_l^2(\vec p,0)=\sigma_1^2(\vec p,0)=\sigma_2^2(\vec p,0)=\sigma^2.$$
This equation makes clear the physical meaning of the parameter $\sigma$ in
Eq. (\ref{eq7}).
\begin{quote}
\underline{Note.} The quantities $\sigma_j^2(t)$ occurred in
paper \cite{c13}, see Eq. (24) in App.~A. There they played the role of
notations, their physical meaning being not revealed. It was shown here
that these quantities (denoted as $\sigma_j^2(\vec p,t)$ here) do have
the meaning of packet dispersions at the moment $t$. Remark also the error
in writing the expression for $\sigma_3^2(t)$ in Eq. (24) in \cite{c11}:
there $(1-v^2)$ must be squared, cf. Eq. (\ref{eq22}) here.
\end{quote}


\section{Discussion}
\label{s3}

Let us compare the dispersions of the moving packet and the packet at rest.
Using the designations
$$E=\sqrt{p^2+m^2}=\gamma m, \quad \gamma =(1-v^2)^{-1/2},$$
rewrite Eqs. (\ref{eq21}), (\ref{eq22}) in the form
\begin{eqnarray}
\label{eq25}
\sigma_l^2(\vec p,t)&=&\sigma^2+t^2/\gamma^6m^2\sigma^2,\\
\label{eq26}
\sigma_1^2(\vec p,t)&=&\sigma_2^2(\vec p,t)
=\sigma^2+t^2/\gamma^2m^2\sigma^2.
\end{eqnarray}
Comparing with the dispersions $\sigma^2(0,t)$ of the packet at rest,
see Eq. (\ref{eq24}), one obtains
\begin{eqnarray}
\label{eq27}
\sigma_l^2(\vec p,t)&=&\sigma^2(0,t/\gamma^3),\\
\label{eq28}
\sigma_1^2(\vec p,t)&=&\sigma_2^2(\vec p,t)=\sigma^2(0,t/\gamma).
\end{eqnarray}
The transversal dispersions of the moving packet evolves slower than the dispersions
of the packet at rest. Its slowing down is Einsteinian: the dispersions
$\sigma_1^2(\vec p,t)$ and $\sigma_2^2(\vec p,t)$ at the moment $t$ assume the
value which $\sigma^2(0,t)$ assumes at the earlier moment $1/\gamma$ (see ER
definition in the Introduction).

The longitudinal spreading $\sigma_l^2$ also grows slower than $\sigma^2(0,t)$,
but the retardation of $\sigma_l^2$ is not ER, see Eq. (\ref{eq27}). So ER fails.
I suppose that this result deserves its detailed derivation in Sect.~\ref{s2}.
The derivation used the following premises.

The packet of scalar (spinless) particle is described by the wave function
$\Psi (\vec x,t)$ which satisfies relativistic positive-energy Klein-Gordon equation.

The initial packet state is described by the simple Gaussian function of macroscopical
space size.

To calculate $\Psi (\vec x,t)$, the usual approximation was exploited, see Eq. (\ref{eq12}).

Using $\Psi (\vec x,t)$ the packet dispersions at the moment $t$ were obtained.
Unlike $\Psi (\vec x,t)$ the dispersions are the observable nonstationary quantities
which can be experimentally measured.

Examples of nonstationary systems for which ER fails were given in \cite{c5}-\cite{c7}.
The considered
system complements the examples. So one may state that ER is not a universal (kinematic)
law for quantum clocks.

However, in the case of unstable particles quantum mechanics shows that ER holds with
high precision \cite{c3}-\cite{c5}. Experiments also agree with ER, see e.g. the corresponding
references in \cite{c3}-\cite{c6}. As was argued in Introduction, usual explanation of ER
(based on Lorentzian
transformations of position and time) is nonapplicable for quantum clocks. It is
quantum mechanics which provides the suitable theoretical explanation.




\end{document}